# Variation of the bonding interactions and magnetism in GdAu$X$ ($X$ = Mg, Cd, and In)

Hem Chandra Kandpal[1], Frederick Casper[1], Gerhard H. Fecher[1], Claudia Felser[1]
Jürgen Kübler[2] and Rainer Pöttgen[3]

[1] *Institut für Anorganische Chemie und Analytische Chemie, Johannes-Gutenberg-Universität Mainz, Staudinger Weg 9, D-55099 Mainz, Germany*

[2] *Institut für Festkörperphysik, Technische Universität Darmstadt, Hochschulstrasse 6, D-64289 Darmstadt, Germany*

[3] *Institut für Anorganische und Analytische Chemie, Westfälische Wilhelms-Universität Münster, Corrensstrasse 30, D-48149 Münster, Germany*

E-mail: felser@uni-mainz.de; FAX: +49–6131–39–26267

**Abstract**

Results of first-principles electronic structure calculations for the isotypic compounds GdAu$X$ (ZrNiAl type, $X$ = Mg, Cd, and In) are presented. We report on a systematic examination of the electronic structure and nature of the bonding in these intermetallics. Our calculations indicate a metallic state for all of the compounds. We find that the indium in GdAuIn and magnesium in GdAuMg have significant bonding interactions with Au. We have also identified In $s$ lone pair in GdAuIn has more localized behaviour as compared with Mg $s$ in GdAuMg. The magnetic properties are well described within the local density approximation.

## 1. Introduction

Intermetallic gadolinium compounds are promising candidates for magnetocaloric materials (magnetocaloric effect; MCE) [1–2] and magnetoelectronics [3]. The recent developments in this field of magnetocalorics have been reviewed earlier [4]. Although gadolinium and gadolinium-based solid solution alloys are the excellent candidates for MCE materials, some intermetallic gadolinium compounds show adiabatic temperatures that are up to 30 % higher than those for elemental gadolinium. A highly interesting compound in that respect is the giant-MCE material $Gd_5Ge_2Si_2$ [5]. Another class of materials concerned the $Fe_2P$ related pnictide solid solution $MnFeP_{0.5}As_{0.5-x}Ge_x$ [6].

We have recently started a more systematic investigation of the structure-property relations of Gadolinium compounds GdAuZ (Z=Sn, In, Cd, Mg). GdAuSn crystallizes like the many other 18 electron compounds assuming that the *f*-electrons are localized in the LiGaGe/ NdPtSb Structure ($P6_3mc$) [7,8]. The other three compounds obey the $Fe_2P$ / ZrNiAl structure. The isotypic compounds GdAuMg [9], GdAuCd [10], and GdAuIn [11] are antiferromagnetic with distinctly different magnetic ordering temperatures of $T_N$ = 81.1 K, $T_N$ = 66.5 K, and $T_N$ = 12.5 K, respectively. They have similarity in their crystal structure despite the different bonding behaviour of Mg, Cd and In. Mg is more electropositive as compared to Cd and In. GdAuMg and GdAuCd (16 and 7*f*-electrons) have the same number of valence electrons, whereas the indium compound exhibits 17 valence electrons. There is not much known about the structure – property relationship of these compounds. More recently Tjeng *et al* have systematically studied the band structure of GdAuMg within local density approximation LDA and LDA+*U* with a view to understanding the role of partial density of states near the Fermi energy [12]. They have found Gd in GdAuMg is in a half filled $4f^7$ configuration and the states in the vicinity of Fermi energy are unaffected by the Gd 4*f* states.

In this paper, we report the results of spin polarized, tight binding electronic structure calculations performed using the linearised muffin-tin orbital (LMTO) method within the atomic sphere and local density approximations. The LMTO method is appropriate to describe the bonding interaction of the valence electrons in these compounds. For comparison we also performed linearized augmented plane-wave (LAPW) calculations. As we will remark, there are striking similarities in the electronic structures of these compounds, but also remarkable differences. The comparison also brings out the role of s lone pair of In atoms, since in the three compounds, the nature of the s lone pair are different. To this end, we present an analysis of the crystal orbital Hamiltonian populations (COHP), a recently developed tool for the analysis of specific bonding between atoms [13].

## 2. Crystal structures and details of the calculations

The intermetallic compounds GdAuMg, GdAuCd, and GdAuIn crystallize in hexagonal ZrNiAl type structure. In figure 1, a projection of the GdAuMg structure and the corresponding coordination polyhedra are presented as examples. Both crystallographically independent gold sites have a trigonal prismatic coordination by gadolinium and $X$ atoms, respectively. The two different trigonal prismatic building groups are shifted with respect to each other via half the translation period $c$. The trigonal prisms are capped by three additional atoms on the rectangular faces leading to a coordination number 9, which is often observed for related intermetallics. Since the crystal chemistry of these compounds has been discussed in detail in the previous reports on the magnetic and $^{155}$Gd Mössbauer spectroscopic studies [9–11], we only present a brief account here.

The largest differences between the GdAuMg, GdAuCd and GdAuIn structures are the Au-$X$ distances (table 1). There are two in-equivalent Au atoms in the cell. One Au atom has three (in plane) and the other has six nearest $X$ neighbours. Each has two different set of distances for two inequivalent Au atoms. In all three compounds the Au-$X$ distances cover the range from 277 to 291 pm.

All calculations refer to the GdAu$X$ ($X$ = Mg, Cd and In) compounds, ZrNiAl type, of space group $P\bar{6}2m$. The experimental structural parameters that were used as starting values for the calculations are displayed in table 2. The electronic structures presented here were calculated using the self-consistent, scalar relativistic linearized muffin-tin orbital (LMTO) calculations within the local spin density approximation (LSDA), as implemented in the STUTTGART TB-LMTO-ASA program [14]. We have also performed linearized augmented plane-wave (LAPW) calculations based on the Wien2k code [15]. Spin-orbit coupling was however ignored. 462 irreducible *k*-points within the primitive wedge of the Brillouin zone were employed. The default atomic basis sets were used for all the atoms, along with the so-called downfolding procedure [16] applied to certain orbitals. In the LMTO-ASA procedure, the space of the unit cell is filled using both atomic spheres as well as empty spheres whose centers and radii are determined automatically. The empty spheres were described using a 1*s* orbital basis with 2*p* downfolding. In order to avoid unphysical COHP interactions between empty spheres and atoms, all empty sphere orbitals were kept downfolded for COHP calculations. Wien2k is used to overcome the empty sphere problem of LMTO.

To study the electronic structure of antiferromagnetic GdAu*X* (*X* = Mg, Cd and In), a periodic supercell has been used. This can be constructed by doubling the lattice parameters in all three directions. Where Gd atoms lying in one plane are aligned parallel to each others and the next nearest neighbour Gd atoms sitting at other planes are aligned in opposite direction. Such arrangement of spin of Gd atoms avoids the possibility of spin frustration state within the triangular network of Gd atoms.

## 3. Results and Discussion

We calculated the total energies of GdAu*X* (*X* = Mg, Cd and In) in the ferromagnetic and antiferromagnetic states using LMTO and full potential linear augmented plane wave scheme within Wien2k. The differences between total energies for ferromagnetic and the antiferromagnetic GdAu*X* are presented in table 3. The differences in the energies are small in both the calculations. In fact Wien2k does not give the correct ground state found experimentally and therefore a competition between the two ground states can not be excluded. Only LMTO gives correct ground state for GdAuMg. When compared to the other compounds, it is to be noted that GdAuMg clearly exhibits the strongest antiferromagnetic behaviour.

### *3.1. Density of states*

In this section, we present the density of states (DOS) for three GdAu*X* (*X* = Mg, Cd and In) compounds. The DOS for each is displayed in the panels of figure 2 for the different *X* atoms with orbital projected d and f states of Gd. In every one of the three compounds, the DOS exhibits a narrow band composed largely of Gd 4*f* orbitals clustered at approximately –4.3 eV below the Fermi energy (taken as the top of the valence band and set to zero on the energy axis in all the plots). Our density of states indicating that the Gd has localized nature of f electrons and half filled $4f^7$ state as was found from experiment and previous calculations [12, 17]. The authors are aware that the LSDA underestimate the electron correlation in *f* systems, but from reference [12] it is known that the Gd *f* states do not affect the valence states, the Au *d* and In/Cd/Mg *s* and *p* states.

The orbital projected density of states of the valence electrons is shown in the figure 3. In GdAuMg, the density of states shows a band composed largely of Au 5*d* and 6*s* orbitals and a small contribution from Mg 3s 3*p* orbitals ranging from -7 eV to -4 eV below the Fermi energy, while most of the Mg *s* and *p* states are found around the Fermi energy. The metallicity of the compound is mainly due to Mg *s* and *p* states, which appear at the Fermi energy.

Remarkable is the fact that in the In compound, there are small *d* states found at -7 eV close to the In *s* states (see in figure 3(e) and (f)). Some of In *p* states are found with the Au *d* states over a range of -6.5 to -4 eV below Fermi energy. In *s* states are less disperse in GdAuIn compared with Mg *s* states in GdAuMg. Reflecting the smaller dispersion of the In *s* states are correspondingly fewer Au *d* states in this region of energy (see figure 3), -7 eV with respect to the Fermi energy. One observe more pure In *s* states in the In compound, whereas in the Mg case, some of the Mg *s* and Au *d* states are found at the same energies. There is a separation between In *s* and *p* states in GdAuIn compound. Despite of these separation between the *s* and *p* states of In in GdAuIn compared to Mg *s p* in GdAuMg, it is In that has more localized (in energy) *s* states as seen from the narrower DOS of In 5*s* compared with Mg 3*s*.

In the Cd compound, the *d* states of two inequivalent Au atoms are not approximately at the same energy like Mg and In compounds which makes two narrower *d* states. The two inequivalent Au atoms in GdAuCd have Cd atoms in the surroundings, one has six Cd atoms in different planes and the second has three Cd in the same plane. Cd atom has complete filled shell structure. Therefore the Au atom which sits near to the Cd surroundings (not in the same plane) feels more repulsion which causes the shift of the Au *d* states. Due to which we see two more like pure Au *d* states in Cd compound. Cd 4*d* states are centered at -9 eV with some Au 5*d* states. Cd 5*s* and 5*p* states are appear close to the Fremi energy.

The most obvious difference in the DOS of the two compounds (GdAuMg and GdAuIn) is the presence of filled 5*s* orbitals in indium, indicated by a narrow peak at −7 eV below the Fermi energy. The 5*d* states of gold in the magnesium and indium compounds are distinctly broader than those of the cadmium compound. In the magnesium and indium compounds, the 5*d* states of the gold atoms trace the *s p* states of magnesium and *p* states of indium. One observes more pure Au 5*d* states in the cadmium compound. Gd 5*d* states and *X s p* states are mainly contributing at the Fermi energy.

*3.2. Chemical Bonding*

The nature of interaction is better explored by plotting the COHPs of Au-*X* interactions. The COHP can be plotted as a function of the energy and can demarcate different bonding, non-bonding and antibonding contributions for specific pairwise interactions.

For the three compounds, we have calculated LMTO COHPs for the Au–*X* interactions. All COHPs were then scaled by the number of interactions in the unit cell. The Au–*X* COHPs of GdAu*X* for the different *X* are displayed in figure 4. For the convention that we use, a positive COHP represents bonding interactions while a negative COHP represents antibonding interactions.

There are bonding states at the Fermi energy for all the compounds confirming the metallic behaviour. These compounds are electronically very stable as seen from a complete absence of any antibonding interaction below the top of the valence band. An interesting feature is the presence of antibonding component in In compound just above the Fermi energy, which is not seen in GdAuMg and GdAuCd compounds. This indicates the In *p* strongly hybridizes with Au states near the Fermi energy. It is clear from the interatomic distances (table 1) that the GdAuIn has almost two equidistant In atoms around the Au atoms and are shorter than the Au-*X* distances in Mg and Cd compounds.

In GdAuCd, There are antibonding states in the COHP below the Fermi energy, centered around -9 eV. When two closed-shell system interact, one expect the number of filled bonding states should be exactly compensated by the number of filled antibonding states below the Fermi energy [18,19]. The presence of antibonding states below the Fermi energy can be inferred as arising from the filled *d* orbitals of Cd and filled *d* orbitals of Au.

The extra electrons in the In-*p* orbital enhance the extent of the favourable bonding hybridization in GdAuIn by pushing down the *s* electron pair from the Fermi energy and making them more localized. The Au–In COHP is more disperse in GdAuIn indicating Au–In hybridization over a broader energy range. In particular, the COHP strength in the region of the In-*s* states is increased significantly.

We can say the presence of the bonding Au-*X* states in the COHP of the GdAuMg, as arising due to the lone pair on Mg *s* being degenerate over a number of sites. In the GdAuIn, where the lone pair of In *s* is more localized this interaction is no longer possible up to the extent like Mg-Au in GdAuMg compound. This suggest that from a chemical viewpoint, the Au-X interactions is perhaps greater for Mg and In than for Cd.

The dashed lines in the figure 4 are an integration of the COHP up to the Fermi energy, yielding a number that is indicative of the strength of the bonding. The extent of the bonding of Mg and In are different with Au2, but so is the value of the integrated COHP: 1.229 eV per interaction for Mg and 1.33 eV per interaction for In compounds.

The gold and *X* atoms together build up rigid three-dimensional [Au*X*] networks in which the gadolinium atoms fill distorted hexagonal channels. The Au–In (282–290 pm), Au–Mg (278–291 pm) and Au–Cd (280–290 pm) distances compare well with the sums of the covalent single bond radii of 283 pm (Au + In), 270 pm (Au + Mg), and 275 pm (Au + Cd) [20]. For GdAuIn, the shorter Au–In distances match perfectly with the sum of the radii, while they are slightly longer for GdAuMg and GdAuCd. From this comparison we can assume that there are slightly stronger Au–In interactions in GdAuIn than in GdAuMg and GdAuCd, which can be seen from the Au-*X* COHP calculations.

In GdAuMg, the COHPs between gadolinium and magnesium are much weaker as a result of the longer Gd–Mg distances. The extents of such interactions are small but noticeable in the magnesium and indium compounds.

*4. Conclusions*

We have investigated the electronic structure of the isotypic GdAuX (X = Mg, Cd and In) compounds. In GdAuMg and GdAuIn, we find that the Au *d* states are broader than those observed in the corresponding cadmium compound. However, the valence band states in the magnesium and indium compounds have a strong Au *d* admixture. In the case of the cadmium compound, states below the top of the valence band have a more pure Au *d* character.

In particular, we examined the effect of Au-*X* covalency in these compounds. The Au *d* states occupy the same energy range as the In and Mg states, suggesting a significant Au-*X* covalency. We have also identified In *s* lone pair in GdAuIn has more localized behaviour as compared with Mg *s* in GdAuMg. Our electronic structure calculations predict metallic behaviour with the appearance of Gd *d*, Mg *s p*, In *p* and Au *s* states at the Fermi energy.

**Acknowledgements**

This work was supported by the Deutsche Forschungsgemeinschaft through SPP 1166 *Lanthanoidspezifische Funktionalitäten in Molekül und Material*.

**References**

[1]   Pecharsky V K and Gschneidner Jr. K A 1997 *Appl. Phys. Lett.* **70** 3299

[2]   Pecharsky V K and Gschneidner Jr. K A 1998 *Adv. Cryog. Eng.* **43** 1729

[3]   Gschneidner Jr. K A and Pecharsky V K, 2000 *Annu. Rev. Mater. Sci.* 30 387

[4]   Gschneidner Jr. K A, Pecharsky V K and Tsokol A O 2005 *Rep. Prog. Phys.* **68** 1479

[5]   Pecharsky V K and Gschneidner Jr. K A 1997 *Phys. Rev. Lett.* **78** 4494

[6]   Brück E, Tegus O, Zhang L, Li X W, de Boer F R and Buschow K H J 2004 *J. Alloys Compd.* **383** 32

[7]   Sebastian C P, Pöttgen R 2006 *Z. f. Naturf. B: Chemical Sciences*  **61(8)**  1045

[8]   Casper F, Ksenofontov V, Kandpal H C, Reiman S, Shishido T, Takahashi M, Takeda M, Felser C 2006 *Z. f. Anorg.Allg. Chem.* **632(7)**, 1273.

[9]   Łątka K, Kmieć R, Pacyna A W, Fickenscher Th, Hoffmann R -D and Pöttgen R, 2004 *Solid State Sci.* **6** 301

[10]  Łątka K, Kmieć R, Pacyna A W, Fickenscher Th, Hoffmann R -D and Pöttgen R 2004 *J. Magn. Magn. Mater*. **280** 90

[11]  Pöttgen R, Kotzyba G, Görlich E A, Łątka K and Dronskowski R 1998 *J. Solid State Chem.* **141** 352

[12]  Gegner J, Koethe T C, Wu H, Hu Z, Hartmann H, Lorenz T, Fickenscher T, Pöttgen R and Tjeng L H 2006 *Phys. Rev. B* **74** 073102

[13]  Dronskowski R and Blöchl P E 1993 *J. Phys. Chem.* **97** 8617

[14]  Jepsen O and Andersen O K 2000 TB-LMTO-ASA-47 MPI für Festkörperforschung Stuttgart Germany

[15]  Blaha P, Schwarz K, Madsen G K H, Kvasnicka D and Luitz J 2001 *WIEN2k, An Augmented Plane wave + Local Orbitals Program for Calculating Crystal Properties*, Karlheinz Schwarz, (Techn. Universität Wien, Austria)

[16]  Jepsen O and Andersen O K 1995 *Z. Phys*. B **97** 35

[17]  Casper F, Kandpal H C, Fecher G H and Felser C *J Phys.D: Appl. Phys.* 2007  **40** 3024

[18]  Seshadri R and Hill N A 2001 *Chem. Mater.* **13** 2892

[19]  Raulot J-M, Baldinozzi G, Seshadri R and Cortona P 02 *Solid State Sci.* **4** 467

[20] Emsley J 1989 *The Elements* Clarendon Press Oxford

**Table 1.** Interatomic distances (pm) in the structures of GdAu$X$ ($X$ = In, Mg and Cd) [7–9]. All distances within the first coordination sphere are listed

|      |   |     | GdAuMg | GdAuCd | GdAuIn |
|------|---|-----|--------|--------|--------|
| Gd:  | 4 | Au1 | 307.6  | 306.4  | 306.8  |
|      | 1 | Au2 | 312.0  | 312.4  | 312.8  |
|      | 2 | $X$ | 330.6  | 326.1  | 325.5  |
|      | 4 | $X$ | 341.3  | 338.0  | 338.8  |
|      | 4 | Gd  | 395.1  | 396.0  | 404.6  |
|      | 2 | Gd  | 412.7  | 405.1  | 397.8  |
| Au1: | 3 | $X$ | 290.8  | 290.2  | 289.5  |
|      | 6 | Gd  | 307.6  | 306.4  | 306.8  |
| Au2: | 6 | $X$ | 277.8  | 280.4  | 281.6  |
|      | 3 | Gd  | 312.0  | 312.4  | 312.8  |
| $X$: | 2 | Au2 | 277.8  | 280.4  | 281.6  |
|      | 2 | Au1 | 290.8  | 290.2  | 289.5  |
|      | 2 | $X$ | 322.2  | 326.1  | 345.2  |
|      | 2 | Gd  | 330.6  | 338.0  | 325.5  |
|      | 4 | Gd  | 341.3  | 344.0  | 338.8  |

**Table 2.** Experimental crystal structures of GdAuX (X = Mg, Cd and In), space group P$\bar{6}$2m (No. 189); Gd in (x, 0, 0); Au1 in (1/3, 2/3, 1/2); Au2 in (0, 0, 0); X in (x, 0, 1/2)

| Compounds | a (Å) | b (Å) | x (Gd) | x ($X$) | Reference |
|-----------|-------|-------|--------|---------|-----------|
| GdAuMg | 7.563 | 4.1271 | 0.41250 | 0.7540 | [7] |
| GdAuCd | 7.701 | 3.960 | 0.4057 | 0.7421 | [8] |
| GdAuIn | 7.698 | 3.978 | 0.40635 | 0.7411 | [9] |

**Table 3.** Total energy difference $\Delta E$ (in eV) between the ferromagnetic (FM) and antiferromagnetic (AFM) structures of GdAu$X$ ($X$ = Mg, Cd and In) and corresponding Néel temperature in Kelvin.

| Compounds | $\Delta E$ (LMTO) | $\Delta E$ (Wien2k) | $T_N$ |
|---|---|---|---|
| GdAuMg | -0.0941 | 0.0082 | 81.1 |
| GdAuCd | 0.0087 | 0.0100 | 66.5 |
| GdAuIn | 0.0291 | 0.0023 | 12.5 |

**Figure captions**

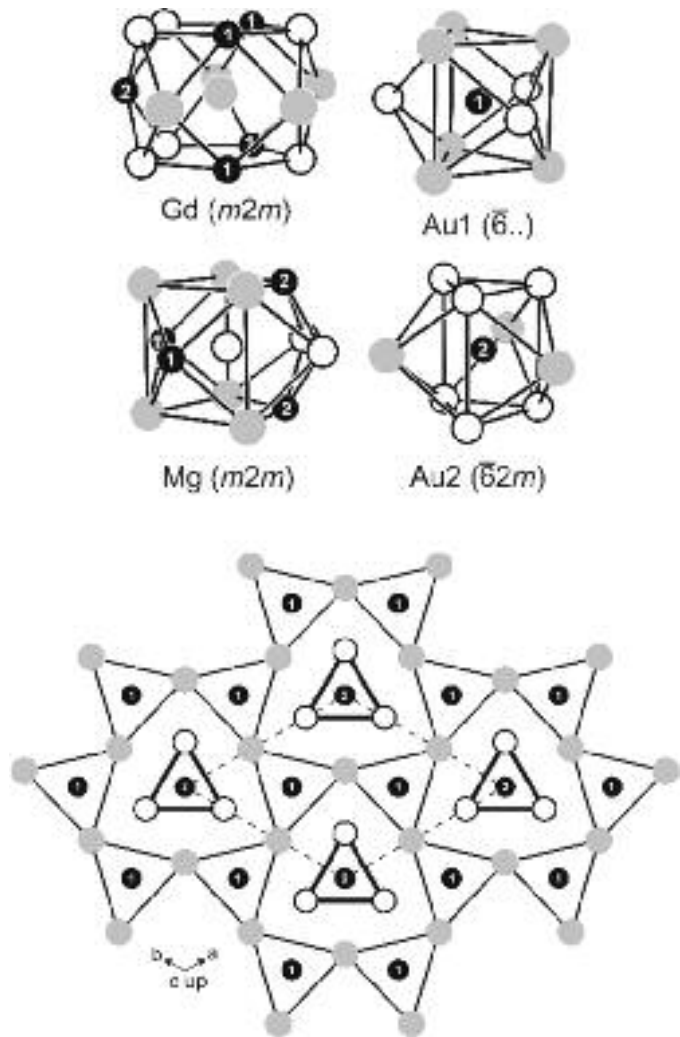

**Figure 1.** Projection of the GdAuMg structure onto the *xy* plane. All atoms lie on mirror planes at *z* = 0 (thin lines) and *z* = 1/2 (thick lines). Gadolinium, gold, and magnesium atoms are drawn as grey, black, and open circles, respectively. The trigonal prisms around the gold atoms are emphasized. The coordination polyhedra are drawn in the upper part and the site symmetries are indicated.

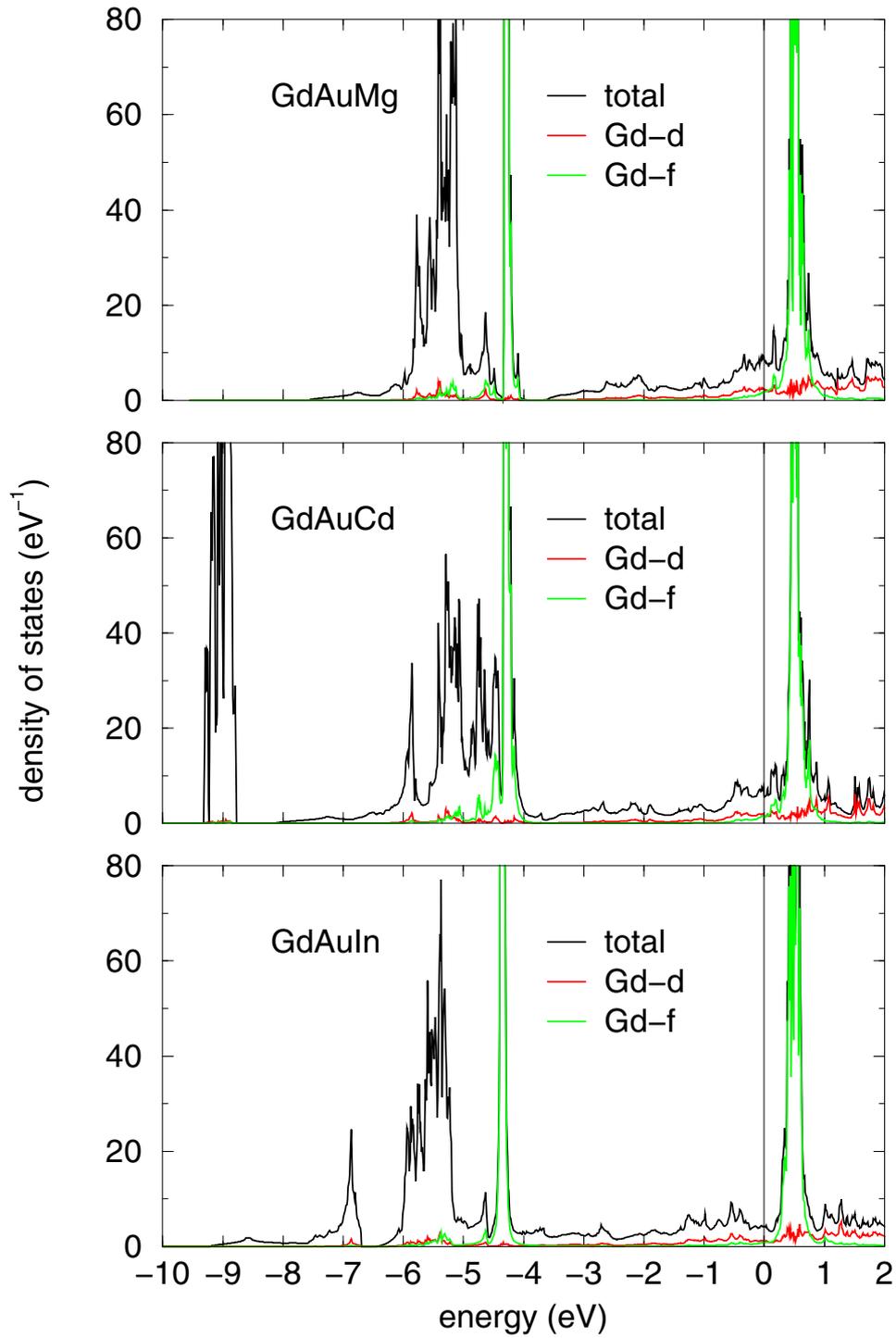

**Figure 2.** LMTO Total DOS and orbital projected Gd f and Gd d DOSs for GdAu*X* (*X* = Mg, Cd and In) compounds. In these and in other plots, the top of the valence band is taken as zero on the energy axis.

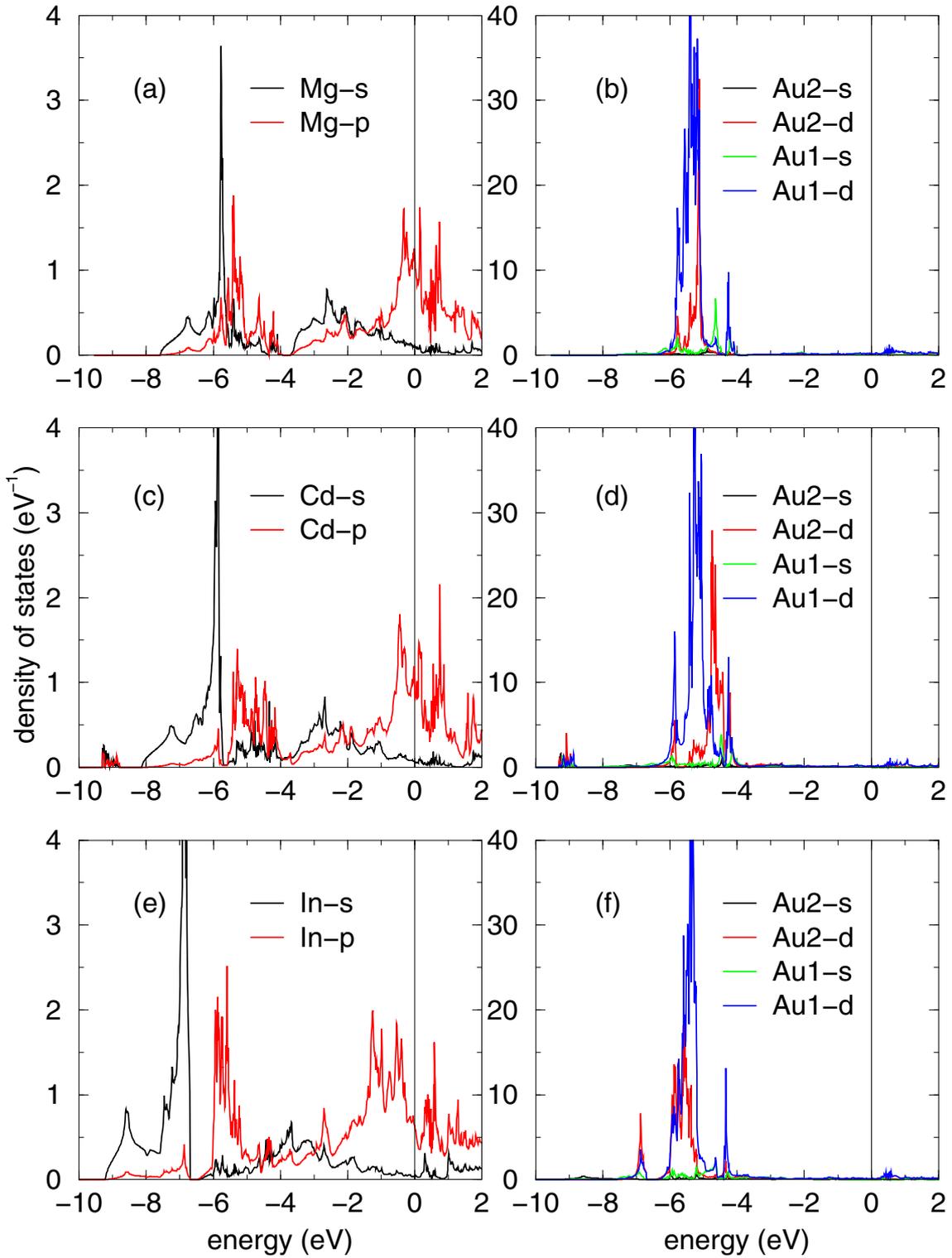

**Figure 3.** Orbital-projected density of states for GdAuMg (a) (b), GdAuCd (c) (d) and GdAuIn (e) (f). The orbitals that contribute to the DOS are labelled.

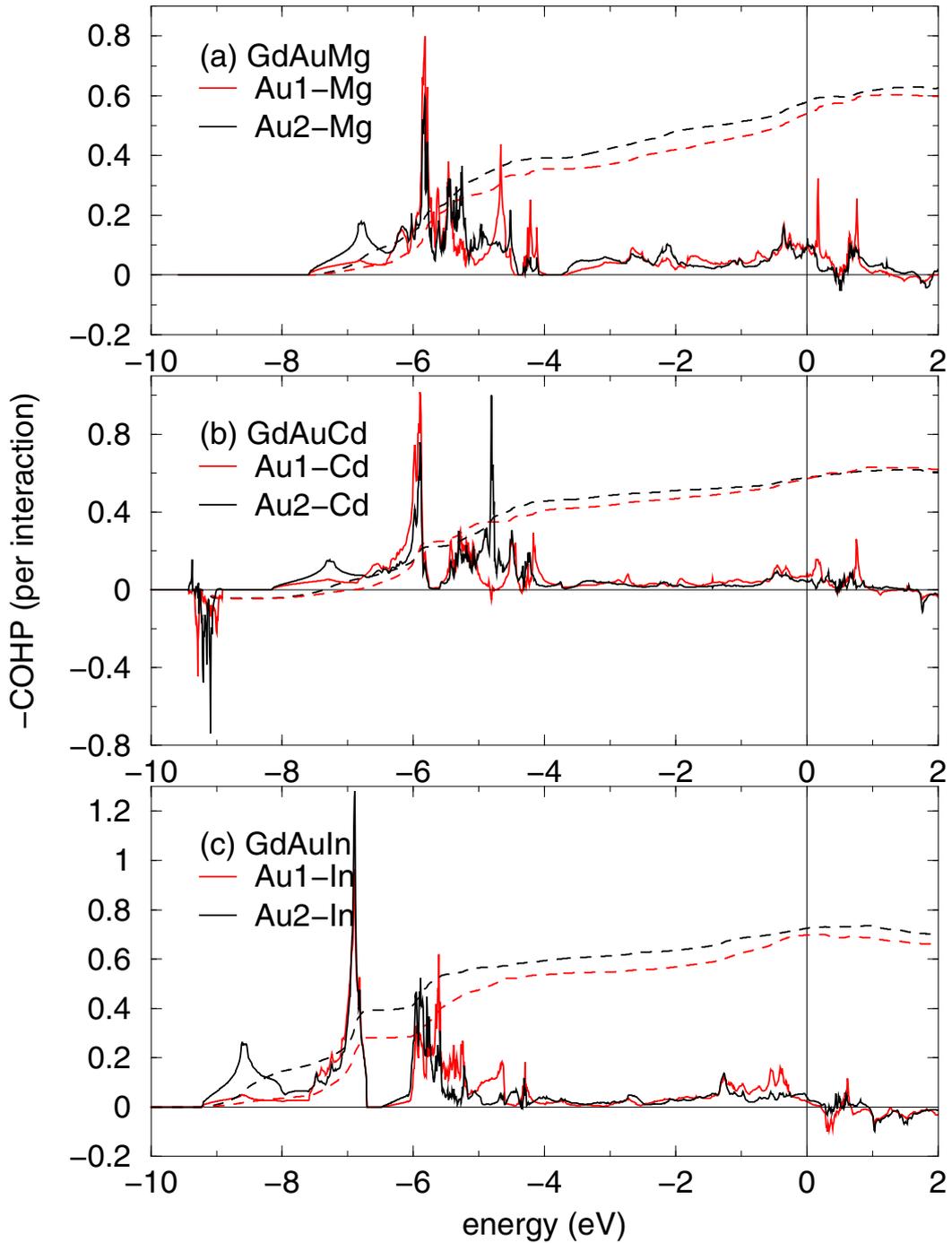

**Figure 4.** Crystal Orbital Hamiltonian Populations (COHPs) for the Au–$X$ interactions in GdAu$X$ ($X$ = Mg, Cd and In). Integrated COHP (ICOHPs) are depicted using broken lines. Note: There are two inequivalent Au atoms in the cell. ICOHPs are obtained by integration of the COHPs from over all valence electrons, i.e., up to the top of the valence band. The values of ICOHPs are (a) Au1-Mg 1.118; Au2-Mg 1.229 (b) Au1-Cd 1.2467; Au2-Cd 1.500 (c) Au1-In 1.08; Au2-In 1.3248 in unit eV per interactions.